# Reproducible Performance Improvements to Monolayer MoS$_2$ Transistors through Exposed Material Forming Gas Annealing

Nicholas B. Guros,[1,2] Son T. Le,[3,4] Siyuan Zhang,[3,4] Brent A. Sperling,[5] Jeffery B. Klauda,[2,*] Curt A. Richter,[4] and Arvind Balijepalli[1,*]

[1]Microsystems and Nanotechnology Division, National Institute of Standards and Technology, Gaithersburg, MD 20899, USA; [2]Department of Chemical and Biomolecular Engineering, University of Maryland, College Park, MD 20742, USA; [3]Theiss Research, La Jolla, CA 92037, USA; [4]Nanoscale Device Characterization Division, National Institute of Standards and Technology, Gaithersburg, MD 20899, USA; [5]Chemical Sciences Division, National Institute of Standards and Technology, Gaithersburg, MD 20899, USA
[*]e-mail: jbklauda@umd.edu, arvind.balijepalli@nist.gov

**Abstract**

Metal-mediated exfoliation has been demonstrated as a promising approach for obtaining large-area flakes of 2D materials to fabricate prototypical nanoelectronics. However, several processing challenges related to organic contamination at the interfaces of the 2D material and the gate oxide must be overcome to realize robust devices with high yield. Here, we demonstrate an optimized process to realize high-performance field-effect transistor (FET) arrays from large-area ($\approx$ 5000 μm$^2$) monolayer MoS$_2$ with a yield of 85 %. A central element of this process is an exposed material forming gas anneal (EM-FGA) that results in uniform FET performance metrics (i.e., field-effect mobilities, threshold voltages, and contact performance). Complementary analytical measurements show that the EM-FGA process reduces deleterious channel doping effects by decreasing organic contamination, while also reducing the prevalence of insulating molybdenum oxide, effectively improving the MoS$_2$-gate oxide interface. The uniform FET performance metrics and high device yield achieved by applying the EM-FGA technique on large-area 2D material flakes will help advance the fabrication of complex 2D nanoelectronics devices and demonstrates the need for improved engineering of the 2D material-gate oxide interface.

Keywords: Field Effect Transistor; MoS$_2$; Forming Gas Annealing; 2D Material Processing; 2D Material Interfaces; Nanoelectronics.

## 1. Introduction

With the scaling of silicon complementary metal-oxide-semiconductor (CMOS) field-effect transistor (FET) technology approaching fundamental limits of device dimensions, power consumption, and heat dissipation,[1-2] an intense effort is underway to develop the next generation of switching devices for use in efficient computation and other low power applications.[3-5] Over the last decade, progress in the use of two-dimensional (2D) materials for numerous applications in the field of nanoelectronics has demonstrated the potential for these materials to transform the semiconductor industry.[6-8] Two-dimensional materials have diverse electronic properties, ranging from semi-metals (e.g., graphene) to semiconductors (e.g., $MoS_2$, $WSe_2$, etc.) to insulators (e.g., hexagonal boron nitride).[9-12] Furthermore, even when these materials are made atomically thin (i.e., a single monolayer), they exhibit good electrical and mechanical properties[13-15] making them ideal candidates for next generation electronics. A broad range of high-performance electronic devices such as FETs,[16-17] light-emitting diodes (LED),[11, 18] photodetectors,[19-20] and biosesnors[21-22] have been realized from 2D materials showcasing their utility in applications where high sensitivity and low power operation are required. However, while the diversity of electronic properties and devices that can be obtained by using 2D materials is virtually limitless, their practical realization is hampered by device fabrication challenges, such as contamination at the interface of the material and gate oxide,[23-25] poor channel doping control,[26-29] and high contact resistance,[26, 30-31] resulting in unreliable device performance.

2D materials can be obtained from either geological sources or through chemical synthesis. Mechanical exfoliation[32] has been traditionally used to obtain 2D materials from geological sources, allowing the fabrication of prototype devices that demonstrate their remarkable properties. However, it is difficult to obtain 2D material flakes with areas large enough to fabricate arrays of nanoelectronics devices or logic circuits using this technique. To overcome this challenge, methods including chemical vapor deposition (CVD)[33-35] and physical vapor deposition (PVD)[36] are being developed to synthesize 2D material flakes with sufficiently large areas. Despite rapid progress in recent years, the performance of devices fabricated from 2D materials generated with these deposition methods have lagged behind the performance of those fabricated from geologically sourced 2D materials.[34, 37-41] In the interim, metal-mediated exfoliation techniques that yield millimeter scale 2D materials[42-44] can permit the realization of large arrays of devices and complex logic circuits. However, 2D material flakes obtained through metal-mediated exfoliation can suffer from both organic and metal contamination originating from multiple adhesive transfer steps, which can degrade device performance through uncontrolled channel doping and charge traps at the 2D material-gate oxide interfaces, making the fabrication of devices with these 2D material flakes difficult.[42, 45-46] Therefore, new processes informed by better characterization of the interface between a 2D material and the gate oxide are needed to improve the performance and reliability of devices fabricated from metal-mediated sourced 2D materials.

We demonstrate a process that improves the performance and reliability of FETs fabricated from $MoS_2$ monolayers obtained by gold-mediated exfoliation.[42] To date, techniques such as ultra-high vacuum (UHV) annealing[47-49] and UV ozone (UV-$O_3$)[48, 50-52] have been applied to multilayer $MoS_2$ flakes obtained with traditional mechanical exfoliation to remove organic contamination. However, their use with monolayers has thus far been avoided because of the risk of destroying the material or generating insulating molybdenum oxide ($MoO_x$). Similarly, forming gas annealing (FGA) has been applied to $MoS_2$ FETs to improve metal-$MoS_2$ contact resistance and also remove organic contamination,[48, 53] but such anneals are usually performed at temperatures between 200°C and 300°C, for short durations (2 – 4 hours), and after the deposition of a top-gate oxide to minimize the risk of material damage and mitigate the creation of sulfur vacancies.[53-54] Forming gas annealing for longer temperature and durations on exposed MoS$_2$ is thought to damage or destroy the material,[54] but we demonstrate it does not.

The processing techniques developed as part of this work, namely an exposed material forming gas anneal (EM-FGA), allow high performance FET arrays to be reliably fabricated from MoS$_2$ obtained from metal-mediated exfoliation. FET performance improvements are a direct result of the EM-FGA improving the 2D material-gate oxide interfaces, which decreases deleterious channel doping without damaging the material, and eliminates the presence of insulating molybdenum oxide MoO$_x$. We show the physical and chemical basis for improved FET performance with complementary analytical measurements using Raman spectroscopy, X-ray photoelectron spectroscopy (XPS), and atomic force microscopy (AFM) to demonstrate the effectiveness of the EM-FGA and its reliability for the fabrication of FETs and potentially other devices fabricated from 2D materials.

## 2. Results and Discussion

### 2.1 Monolayer MoS$_2$ Field-Effect Transistor Fabrication

To realize monolayer MoS$_2$ FETs, MoS$_2$ was first transferred onto an oxidized Si substrate with an oxide (SiO$_2$) thickness of 70 nm using the gold-mediated exfoliation technique described in the *Methods*.[42] Numerous flakes of the transferred material were measured using Raman spectroscopy to have monolayer thickness with an average area between 1000 μm$^2$ and 5000 μm$^2$ as seen from Figure 1a. The Raman peaks corresponding to the E$^1_{2g}$ phonon mode (in-plane vibration for Mo and S at ≈ 386 cm$^{-1}$) and the A$_{1g}$ mode (out-of-plane vibration for Mo and S at ≈ 403 cm$^{-1}$) were found to be in good agreement with the expected shift[55] for monolayer MoS$_2$ (thickness ≈ 0.7 nm) and yielded a frequency difference of 16.6 cm$^{-1}$ (Figure 1b, *orange*).

Furthermore, the frequency difference between the $A_{1g}$ and $E^1_{2g}$ peaks increased to 22.4 cm$^{-1}$ for a bilayer and to 24.8 cm$^{-1}$ for bulk MoS$_2$ flakes (Figure S1) in agreement with literature values.[55]

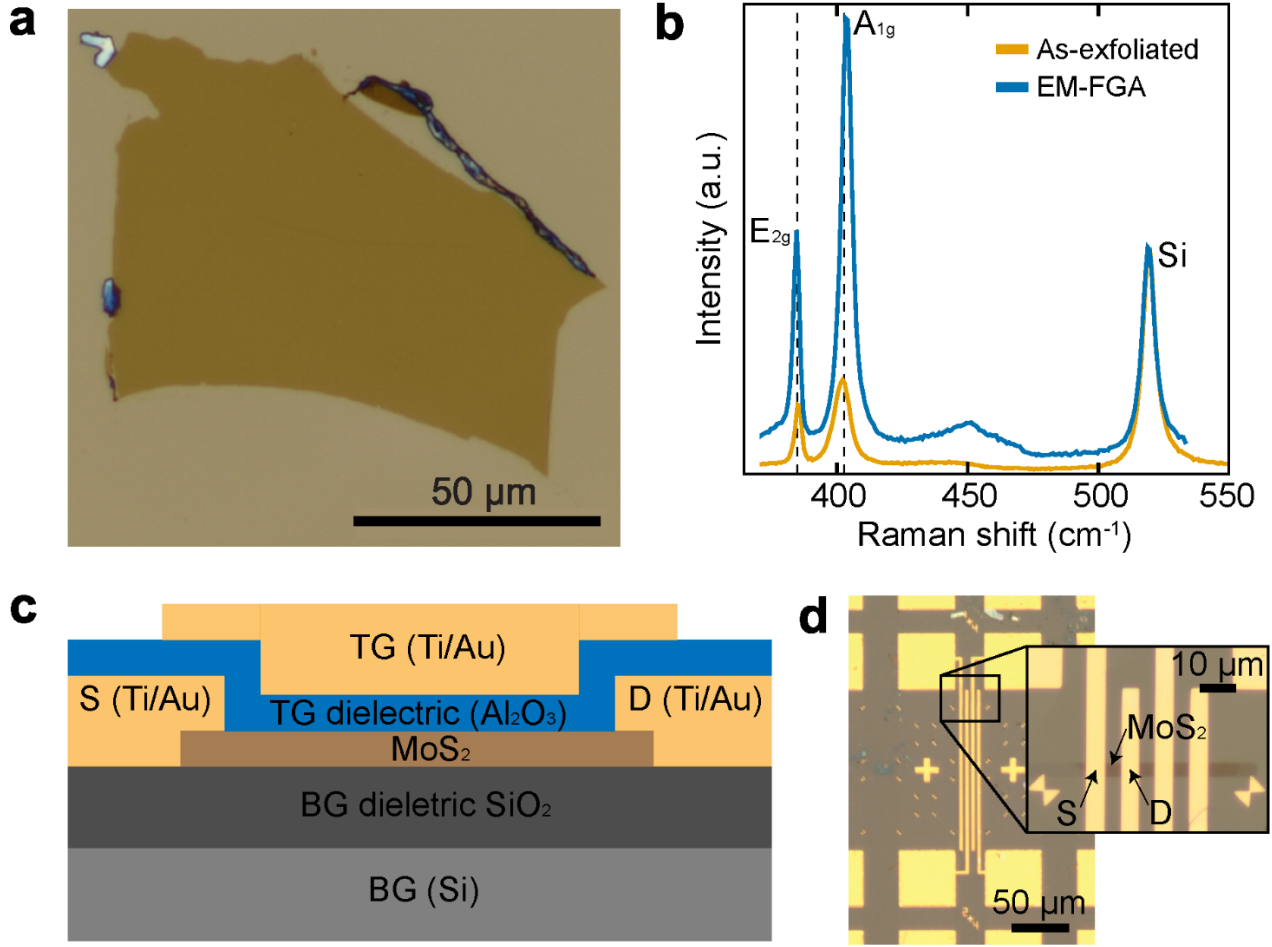

**Figure 1**. MoS$_2$ monolayer characterization and design of monolayer MoS$_2$ field-effect transistors (FET). (a) Large area (≈ 5000 μm$^2$) monolayers of MoS$_2$ were transferred onto a SiO$_2$ on Si wafer using the gold-mediated exfoliation method. (b) Raman spectra of the monolayer from (a) before (*orange*) and after (*blue*) an exposed material forming gas anneal. (c) Cross-sectional schematic depicting a FET fabricated using monolayer MoS$_2$ (550 μm Si back-gate (BG), 70 nm SiO$_2$ BG oxide, monolayer (≈ 0.7 nm) MoS$_2$, 2 nm Ti/100 nm Au sources/drain contacts, 20 nm Al$_2$O$_3$ top-gate (TG) oxide, and 10 nm Ti/100 nm Au TG contact). (d) Optical image of a representative array of FETs prior to deposition of the top-gate dielectric and top-gate metal. *Inset*: Detail view of the FET array.

A schematic of a monolayer MoS$_2$ FET is depicted in Figure 1c (see *Methods* for fabrication details). Briefly, the source (*S*) and drain (*D*) contacts (2 nm Ti/80 nm Au) were patterned by using optical lithography and electron-beam metal deposition after gold-mediated transfer of monolayers. For each FET, a 5 μm × 5 μm channel was lithographically defined and subsequently etched. Figure 1d shows an optical image of an array of three FETs with a global back-gate (*BG*) and back-gate dielectric (*gray,* SiO$_2$). Next, the top-gate (*TG*) dielectric (*blue*, Al$_2$O$_3$) was deposited using atomic layer deposition (ALD) and the top-gate metal (10 nm Ti/100 nm Au) was patterned using optical lithography and electron-beam metal deposition. The large areas and relative abundance of exfoliated monolayers on the substrate allowed for batch fabrication of numerous monolayer FET arrays on a 4-inch wafer.

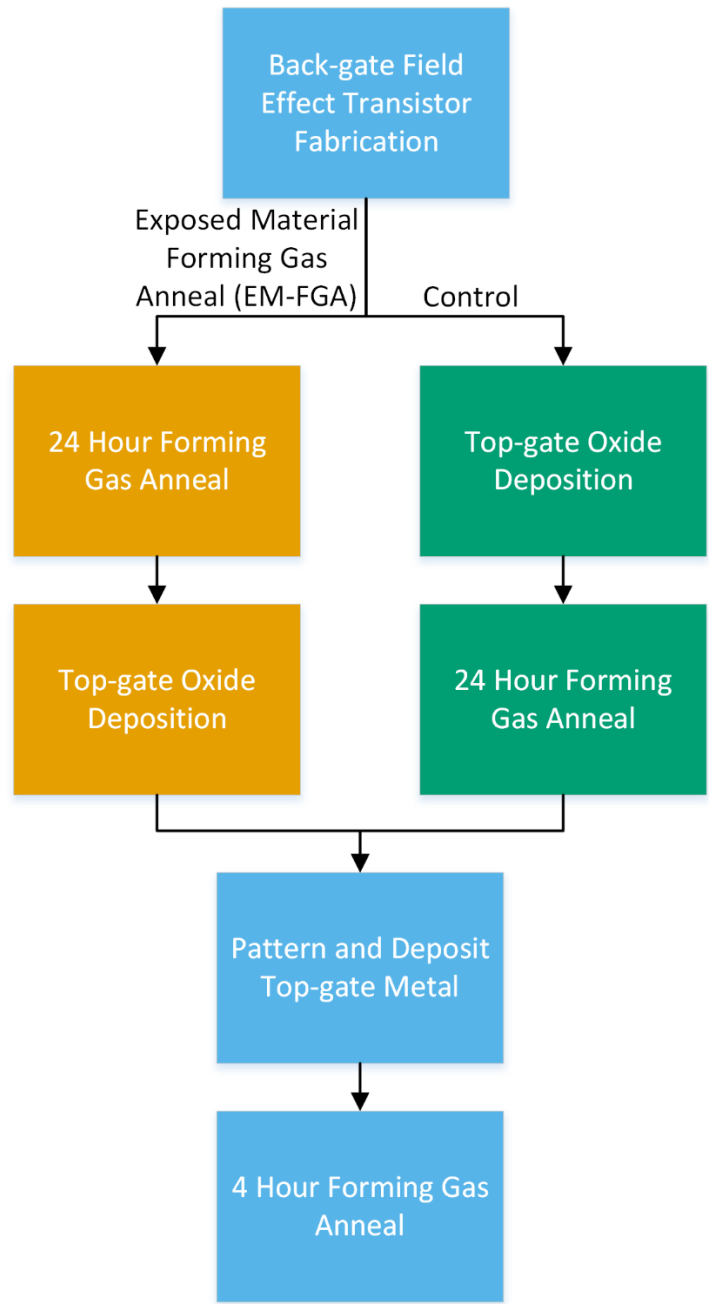

**Figure 2.** Process flow diagram for the fabrication of field-effect transistor (FET) arrays. The *orange* steps (on the left) highlight the newly developed exposed material forming gas anneal (EM-FGA) sequence while the *green* steps (on the right) represent a conventional sequence for FET fabrication from 2D materials. The *blue* steps (top and bottom) are common to both processes.

## 2.2 Forming Gas Anneal Effects on FET Performance

Two sets of FETs were fabricated by using the process flow described in *section 2.1* and shown in Figure 2. A control set (*n*=5) was processed using the steps shown in Figure 2 on the right in *green*, in which a conventional annealing process was used, i.e. the entire set of five control FETs underwent a forming gas anneal (FGA) immediately after deposition of a top-gate oxide.[48, 54] The second set of FETs (*n*=20) was fabricated with our new EM-FGA process as illustrated in Figure 2 on the left in *orange*. We varied the anneal time and gas flow rate (Figure S2) and determined that an anneal temperature of 400 °C with 100 cm$^3$/min forming gas for 24 hours at standard temperature and pressure (STP), 0 °C and 101 kPa, respectively (100 sccm), yielded an optimum improvement in performance. The back-gate performance of both the control and EM-FGA FETs was characterized after deposition of the top-gate oxides, but without the top-gate metals. Next, top-gate metals were deposited onto both sets of FETs followed by a second shorter FGA (Figure 2). Finally, the top-gate performance of all devices was measured while the back-gate was connected to ground.

Both the EM-FGA and control FETs demonstrated measurable improvement in back-gate performance compared to as-exfoliated (before FGA or top-gate oxide deposition) back-gate performance (Figure S3; *blue*). On average, 85 % (*n*=17/20) of EM-FGA FETs showed consistent and improved performance relative to the control samples. Next, we discuss the electrical characteristics of the EM-FGA FETs compared to the control set.

**Back-gate Performance**

EM-FGA FET back-gate performance after top-gate oxide deposition and prior to top-gate metal deposition is shown by the representative *orange* transfer curve in Figure 3a for $V_{DS}$ = 1.05 V (all transfer curves can be seen in Figure S4). The measurements were repeated for multiple (stepped) $V_{DS}$ as seen in Figure 3b, where minimal hysteresis was observed. Average electrical performance parameters for all the measured devices are reported in Table 1. All devices demonstrated *n*-type behavior, consistent with previous observations for MoS$_2$ FETs.[16, 56] Unless otherwise noted, error bars reported in this work represent the standard error.

**Table 1.** Performance parameters for EM-FGA and control FETs reported as means and standard errors. Several of these metrics are labeled as "N/A" because the large flat band shift in $V_T$ for the control FETs precluded an accurate estimation of these metrics without inducing dielectric breakdown in the back-gate or top-gate oxide.

| Parameter | EM-FGA (n=17) | Control (n=5) |
|---|---|---|
| *Back-gate* | | |
| $\mu_{FE}$ (cm$^2$/V·s) | 16.1 ± 2.4 | 13.5 ± 3.5 |
| $V_T$ (V) | 2.4 ± 0.9 | −21.1 ± 2.2 |
| $I_{on}/I_{off}$ | 10$^5$ | N/A |
| $I_{on}$ (µA/µm) | > 10 | > 10 |
| Subthreshold swing (V/decade) | 4.6 ± 0.3 | N/A |
| *Top-gate* | | |
| $\mu_{FE}$ (cm$^2$/V·s) | 2.8 ± 0.5 | 4.1 ± 0.3 |
| $V_T$ (V) | −1.8 ± 0.3 | N/A |
| $I_{on}/I_{off}$ | 10$^6$ | N/A |
| $I_{on}$ (µA/µm) | > 10 | > 10 |
| Subthreshold swing (mV/decade) | 650 ± 24 | N/A |

On average, we observed an $I_{on}/I_{off}$ ratio of ≈ 10$^5$, and a subthreshold swing of (4.6 ± 0.3) V/decade for the 70 nm SiO$_2$ back-gate interface. At large positive $V_{BG}$, we observed an $I_{on}$ of at least 10 µA/µm and a field-effect mobility ($\mu_{FE}$), not correcting for source and drain contact resistance, of (16.1 ± 2.4) cm$^2$/V·s that was determined using, $\mu_{FE} = \frac{g_{m,max} L/W}{C_{ox} V_D}$, where $g_{m,max}$ is the peak transconductance, $L$ and $W$ are the length and width of the channel respectively, and $C_{ox}$ is the oxide capacitance per unit area, determined to be 49.3 nF/cm$^2$ for the 70 nm SiO$_2$ back-gate dielectric.[57] To prevent the risk of dielectric breakdown, we limited the range of $V_{BG}$ to ±25 V. The threshold voltage ($V_T$), estimated by extrapolating the point of maximum slope on the transfer curve to the *x*-axis, was found to be (2.4 ± 0.9) V for the EM-FGA FETs. In contrast, the control set exhibited a large and negative shift in $V_T$ of (−21.1 ± 2.2) V, as shown by the representative *green* transfer curve in Figure 3a (all transfer curves can be seen in Figure S5). This shift in $V_T$ is the key improvement in performance for the EM-FGA FETs that separates them from the control FETs.

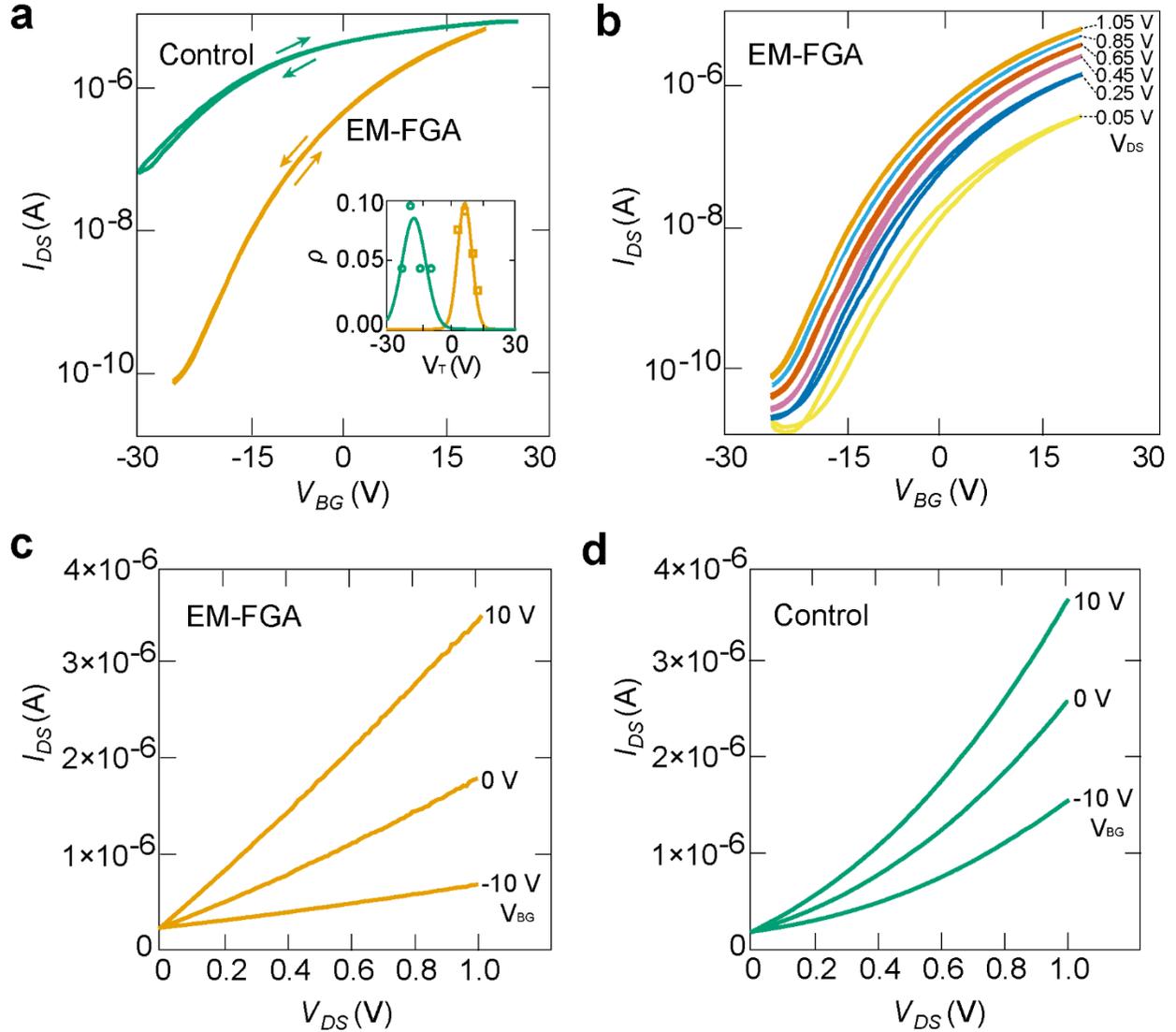

**Figure 3**. Characterization of field-effect transistor (FET) back-gate performance. (a) Representative transfer curves for an exposed material forming gas anneal (EM-FGA) FET (*orange*) and a control FET (*green*) for $V_{DS}$ = 1.05 V. *Inset*: Distribution of $V_T$ for the EM-FGA and control FETs. (b) Representative transfer curves for an EM-FGA FET at varying $V_{DS}$. (c) Representative $I_{DS}V_{DS}$ curves for an EM-FGA FET at varying $V_{BG}$ demonstrate improved contact performance. (d) Representative $I_{DS}V_{DS}$ curves for a control FET at varying $V_{BG}$. All measurements were performed after deposition of a top-gate oxide and prior to the deposition of a top-gate metal.

Notably, a negative shift in $V_T$ of the control set is consistent with previous observations of MoS$_2$ FETs after top-gate oxide deposition.[48, 58] This shift could be explained by the presence of large trapped charges at the MoS$_2$-top gate oxide interfaces that dopes the channel and induces a flatband voltage ($V_{FB}$) shift. To quantify this behavior, we define $V_{FB} = \varphi_{MS} - \frac{Q_i}{C_{ox}}$, where $\varphi_{MS}$ is the difference in the workfunction between the back-gate and the semiconducting MoS$_2$, $Q_i$ is the density of fixed oxide and channel-contaminating charges, and $C_{ox}$ is the back-gate oxide capacitance per unit area. $Q_i$ can be quantified by substituting the definition of $V_{FB}$ into the general gate bias equation, $V_G - V_{FB} = -\frac{Q_s}{C_{ox}} + \psi_s$, where $V_G$ is the gate voltage, $Q_s$ is the charge density of the MoS$_2$ channel, and $\psi_s$ is the surface potential,[59] yielding equation $V_G - \varphi_{MS} \mp \frac{Q_i}{C_{ox}} = -\frac{Q_s}{C_{ox}} + \psi_s$. We can calculate the difference in experimental $Q_i$ for the EM-FGA and control FETs with respect to the ideal case by setting $V_G = V_T$, and assuming several other interface properties ($\varphi_{MS}$, $C_{ox}$, $Q_s$, and $\psi_s$) are the same for both cases. For the ideal case, we assume $Q_i = 0$ yielding $\Delta V_{T(experimental-ideal)} = \frac{-Q_{i,experimental}}{C_{ox}}$.

For a monolayer MoS$_2$ FET, ideal $V_T$ is defined as the $V_G$ at which the quantum capacitance of the channel equals $C_{ox}$.[60] This definition must be used instead of the standard definition of $V_T$, which is only applicable to bulk FETs.[59] Using equations (1) – (3) outlined in *Methods*, the theoretical value of $V_T$ was calculated to be +0.7 V for a FET for a monolayer MoS$_2$ channel on a 70 nm SiO$_2$ oxide. Therefore, the experimentally observed $V_T$ of (2.4 ± 0.9) V for the EM-FGA case compares favorably to the theoretical value. On the other hand, for the control FETs, we measured $V_T = (-21.1 \pm 2.2)$ V, which represents a large and negative shift from ideal $V_T$ (Table 1), indicating the presence of substantial positive contamination that dopes the channel. The preceding results allowed us to estimate $Q_i$ for both the EM-FGA and the control FETs. The estimated value of $Q_i$ is closer to ideal for the EM-FGA FETs ($\approx -4.5 \times 10^{11}$ q/cm$^2$) than for the control FETs ($\approx 6.7 \times 10^{12}$ q/cm$^2$). The order of magnitude reduction in charge, due to the removal of positive contamination, strongly shifts $V_T$ of the EM-FGA FETs in the positive direction and closer to the ideal value of +0.7 V. Furthermore, the estimated value of $V_T$ for the EM-FGA devices is statistically consistent with the ideal value with 95 % confidence. This highlights the importance of the sequence of processing steps developed in this study (see Figure 2) with respect to improving the quality of a 2D material-gate oxide interface.

The benefits of the EM-FGA also extend to improved contact performance in the EM-FGA devices relative to the control set. After the EM-FGA, the $I_{DS}$-$V_{DS}$ response of the FETs as a function of $V_{BG}$ was found to be Ohmic as seen in Figure 3c (all EM-FGA FET $I_{DS}$-$V_{DS}$ responses can be seen in Figure S6 and 2-point resistances can be seen in Table S1). In contrast, Figure 3d demonstrates that the control devices exhibited rectifying characteristics indicating the presence of a Schottky barrier at those contacts (all control FET $I_{DS}$-$V_{DS}$ responses can be seen in Figure S7). We quantified the difference in contact resistance ($R_C$) between the EM-FGA and control FETs using a four-point probe measurement technique (Figure S8) as described in *Methods*. From these measurements, $R_C$ was estimated to be $(35 \pm 3)$ kΩ-µm for the EM-FGA FETs and $(785 \pm 32)$ kΩ-µm for the control FETs, where $R_C$ for the EM-FGA FETs is ten-fold lower than previously reported for monolayer MoS$_2$ FETs.[63]

Forming gas annealing improves contact resistance ($R_C$) between metal source/drain contacts and MoS$_2$ through two mechanisms: 1) by removing organic contamination in the vicinity of the metal contacts, which generates a physical barrier between the metal contacts[49, 53-54] and 2) by locally doping the MoS$_2$ under the source and drain contacts with metal atoms.[63-64] The EM-FGA FETs demonstrate lower $R_C$ compared to the control FETs because the first mechanism is more effective without a top-gate oxide acting as a physical barrier to the removal of organic contamination by hydrogen gas. Furthermore, the second mechanism is more readily permitted in the EM-FGA FETs because organic contamination does not serve as a physical barrier to the doping of MoS$_2$ under the metal contacts with metal atoms. In contrast, the control FETs were annealed after the deposition of the top-gate dielectric, which shields the MoS$_2$-contact metal interface from hydrogen gas penetration, decreasing the effectiveness of organic contamination removal and subsequent doping of MoS$_2$ with metal atoms.

Many of the improvements displayed by the EM-FGA FETs were also observed for the control FETs. For example, we observed minimal hysteresis, while the drive current was found to be at least 10 µA/µm at large positive $V_{BG}$ and $\mu_{FE}$ was (13.5 ± 3.5) cm$^2$/V·s prior to correcting for the contact resistance (Table 1). However, the large shift in $V_T$ for the control FETs precluded an accurate estimation of the $I_{on}/I_{off}$ ratio and the subthreshold swing without inducing dielectric breakdown in the back-gate dielectric.

**Top-gate Performance.** One goal of our approach is to make top-gated monolayer MoS$_2$ FETs for switching or sensing applications. Therefore, after back-gate characterization, the top-gate metal was deposited onto both the EM-FGA and control FETs followed by a second, shorter FGA, to improve top-gate performance. EM-FGA FET top-gate performance is shown by the representative *orange* transfer curve in Figure 4a and reported for all measured devices in Table 1 (all transfer curves can be seen in Figure S9). On average, and similarly to the back-gate results, we measured minimal hysteresis, an $I_{on}/I_{off}$ ratio of ≈ 10$^6$, and a subthreshold swing of (650 ± 24) mV/dec. At large and positive $V_{TG}$, we measured a drive current of at least 10 µA/µm and $\mu_{FE}$ of (2.8 ± 0.5) cm$^2$/V·s before correcting for the contact resistance (and assuming $C_{ox}$ to be 398 nF/cm$^2$ for the top-gate oxide). $V_T$ was found to be (−1.8 ± 0.3) V, estimated by extrapolating the point of maximum slope on the transfer curve to the *x*-axis.

To compare our experimental top-gate $V_T$ of (−1.8 ± 0.3) V to the ideal value, we again used equations (1) – (3) to calculate ideal $V_T$ for a top-gate FET with a monolayer MoS$_2$ channel under a 20 nm Al$_2$O$_3$ oxide. This value was found to be +0.8 V. In contrast to the back-gate performance, the experimentally determined value of $V_T$ does not compare favorably to ideality, indicating that the contaminants doping the channel affect top-gate performance more than back-gate performance. This may be, in part, due to trapping of these fixed charges at the interface degrading gate control. EM-FGA parameters will be further optimized in future work with the aim of reducing $Q_i$ at the MoS$_2$-top-gate oxide interface and to shift top-gate $V_T$ closer to ideality.

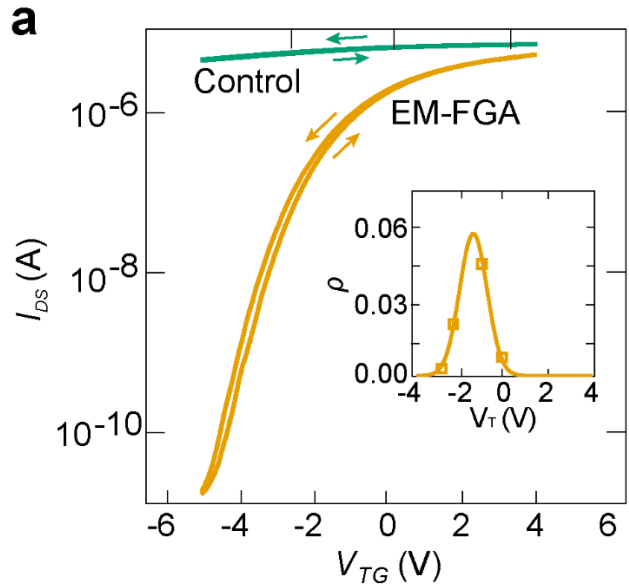
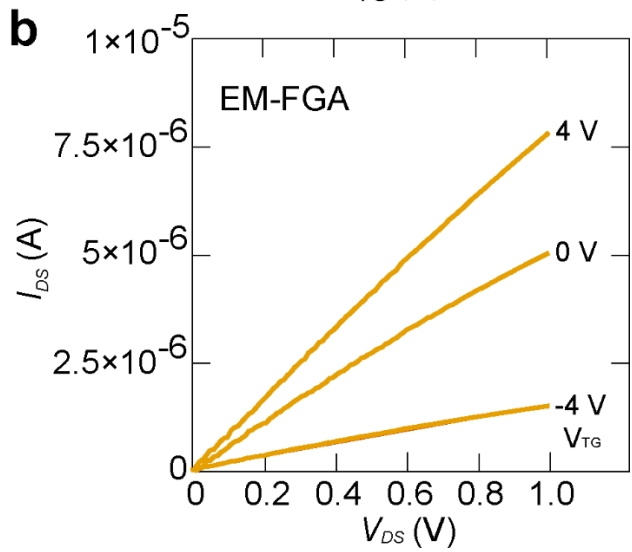
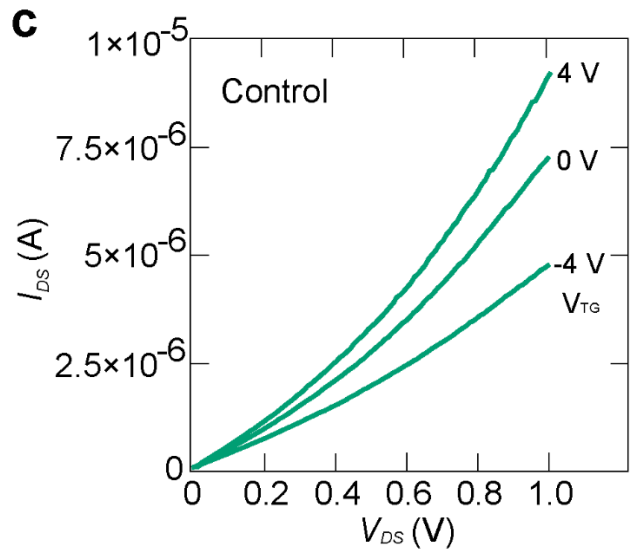

**Figure 4.** Characterization of field-effect transistor (FET) top-gate performance. (a) Representative top-gate transfer curves for an exposed material forming gas anneal (EM-FGA) FET (*orange*) and a control FET (*green*) for $V_{DS}$ = 1.05 V. *Inset*: Distribution of $V_T$ for the EM-FGA FETs (b) Representative $I_{DS}V_{DS}$ curves for an EM-FGA FET at varied $V_{TG}$ demonstrating improvement to contact resistance. (c) Representative $I_{DS}V_{DS}$ curves for a control FET at varied $V_{TG}$. All measurements were made with $V_{BG}$ = 0 V.

Finally, Figure 4b demonstrates that the device $I_{DS}$-$V_{DS}$ characteristics were found to be Ohmic for the EM-FGA FETs (all $I_{DS}$-$V_{DS}$ curves can be seen in Figure S10 and 2-point resistances can be seen in Table S1), in contrast to Figure 4c that demonstrates the rectifying behavior observed for the control FETs (all $I_{DS}$-$V_{DS}$ curves can be seen in Figure S11), similar to the rectifying behavior observed for the control back-gates.

Similarly to the back-gate, some aspects of top-gate performance for the control set were comparable to those of the EM-FGA set. Drive currents approached 10 µA/µm at large and positive $V_{TG}$, and µ$_{FE}$ was found to be (4.1 ± 0.3) cm$^2$/V·s prior to correcting for the contact resistance (Table 1). However, also similar to the back-gate, a large and negative shift in $V_T$ (Figure 4a, *green*) was observed in the top-gate for the control set, which again precluded an accurate estimation of the $I_{on}$/$I_{off}$ ratio, the subthreshold swing, and $V_T$ for the control FETs without inducing dielectric breakdown in the gate dielectric (all transfer curves can be seen in Figure S12). We also note that the large and negative shift in $V_T$ for the control top-gates permitted a more accurate estimate of $g_{m,max}$ than the EM-FGA top-gates where $g_{m,max}$ is likely underestimated because $I_{DS}$ is still increasing at 4 V, which was the maximum $V_{TG}$ that could be applied without inducing dielectric breakdown (Figure 4a). This results in underestimations of $g_{m,max}$ and µ$_{FE}$ for the EM-FGA FETs relative to the control FETs..

In summary, the observed performance benefits of the EM-FGA process are threefold: i) by drastically reducing the interface contamination and trap charges, a controlled $V_T$ shift on both the back-gate and top-gate closer to the ideal value was achieved, which in turn improves performance and reproducibility of FETs fabricated using this approach closer to the level needed for integration in logic circuits,[58, 65] ii) Ohmic metal-MoS$_2$ contacts were achieved, evident by the linear $I_{DS}$-$V_{DS}$ characteristics, with a low contact resistance, and iii) important FET characteristics including $\mu_{FE}$, subthreshold swing, and $I_{on}/I_{off}$ ratio were maintained at values previously reported for FETs fabricated from MoS$_2$ sourced from traditional mechanical exfoliation. Furthermore, the benefits of EM-FGA were achieved using only a tube furnace operating at relatively high pressures, i.e., 350 Pa (2.6 Torr), which makes the technique straightforward to implement without the need for highly specialized equipment.[48-50, 52, 56] The EM-FGA process is gentle and minimizes damage to the large monolayers obtained through metal-mediated exfoliation, unlike other commonly used cleaning techniques that utilize UV-ozone which has been shown to create disadvantageous MoO$_x$ or even eliminate transistor behavior in FETs.[48, 51, 66] We expect EM-FGA to be a critical component of the streamlined processing of 2D materials obtained using increasingly widespread metal-mediated exfoliation techniques.[43, 46, 67-68] Finally, to better describe the mechanism underlying improved FET performance in this work, we performed several complimentary measurements on the monolayers from which the FETs were fabricated, as described next.

## 2.3 Monolayer Characterization with Raman, XPS and AFM

Raman spectroscopy, XPS, and AFM were used to quantify the effects of the EM-FGA on the morphology and chemical composition of MoS$_2$ monolayers obtained from metal-mediated exfoliation. All monolayers analyzed here were prepared using the process steps outlined in *Methods*, identical to the monolayers used to fabricate the FETs, up to the deposition of the top-gate oxide.

**Raman Spectroscopy:** Raman spectra of monolayer MoS$_2$ are shown in Figure 1b. As discussed earlier, the separation between the $E^1_{2g}$ and $A_{1g}$ peaks in the spectrum were in agreement with the expected shift[55] for monolayer MoS$_2$ with a thickness of ≈ 0.7 nm and yielded a frequency difference of 16.6 cm$^{-1}$. The EM-FGA process increased the peak separation frequency to 19.4 cm$^{-1}$, within the range observed for monolayer MoS$_2$.[55] On the other hand, as seen from Figure 1b, samples processed with the EM-FGA demonstrated a dramatic overall increase in the peak intensity (*blue*) relative to the as-exfoliated sample (*orange*), and narrower E$_{2g}$ peak widths (4.9 cm$^{-1}$ compared to 7.3 cm$^{-1}$) suggesting the EM-FGA results in lower contamination and reduces defects in the MoS$_2$ crystal structure. A similar improvement to the material composition was previously observed for multilayer MoS$_2$ annealed with elemental sulfur.[69]

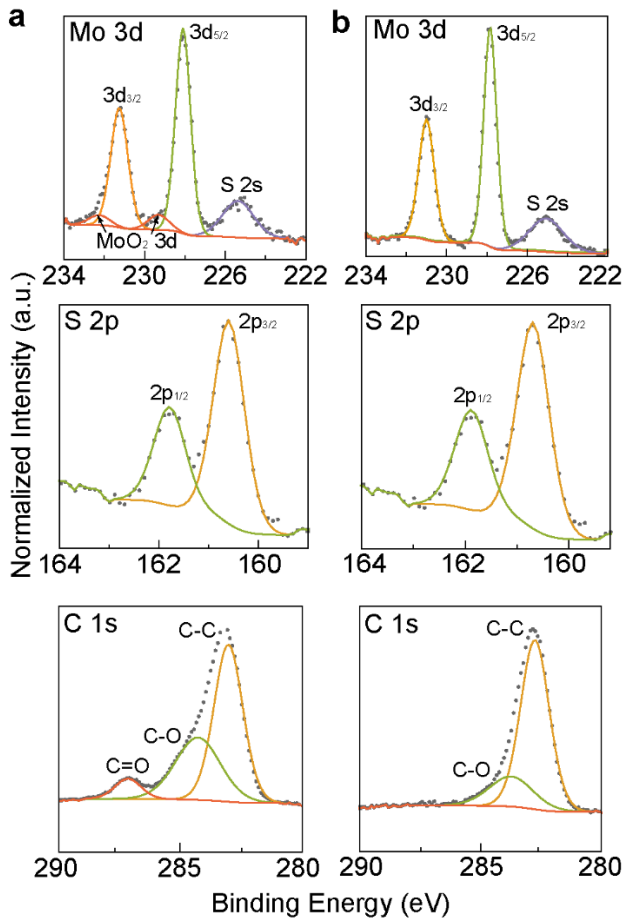

**Figure 5.** XPS spectra for MoS$_2$ sourced from metal-mediated exfoliation before and after exposed material forming gas annealing (EM-FGA). (a) XPS spectra for Mo, S, and C before the EM-FGA and (b) after the EM-FGA illustrating the elimination of MoO$_2$ and species containing C=O bonds, and a reduction of species containing C-O bonds while the presence of Mo$^{4+}$ and S$^{2-}$ were constant.

**X-ray Photoelectron Spectroscopy (XPS):** XPS spectra for three elements, Mo, S, and C, obtained from a MoS$_2$ monolayer before and after EM-FGA is illustrated in Figure 5. All spectral data are calibrated with the C 1s peak at a constant binding energy of ≈ 284.6 eV. Both before and after the EM-FGA, the Mo 3d shows two main peaks at 229.8 eV and 232.9 eV which are attributed to Mo 3d$_{5/2}$ and Mo 3d$_{3/2}$, respectively, confirming the existence of Mo$^{4+}$. The S XPS spectrum displays peaks at 162.7 eV and 163.81 eV that can be attributed to the doublet S 2p$_{3/2}$ and S 2p$_{1/2}$, respectively, corresponding to the divalent sulfur ion (S$^{2-}$) of MoS$_2$. However, the sample measured before the EM-FGA displays ≈ 5 % of MoO$_2$ on the channel, as evidenced by the two peaks at 231.0 eV (Mo 3d$_{5/2}$) and 234.0 eV (Mo 3d$_{3/2}$). These peaks are not observed after the EM-FGA, indicating that EM-FGA removes insulating and disadvantageous MoO$_2$ that forms on the surface of the monolayer.

Figure 5 also demonstrates changes in the C 1s peaks, which illuminate changes in organic contamination. The deconvolution of these peak illustrates the existence of organic compounds before EM-FGA due to the presence of C-C, C-O, and C=O bonds. After the EM-FGA, the intensity of C=O is not detectable and the intensity of C-O decreases, strongly indicating the removal of surface organic contamination. Furthermore, the C content was reduced by more than 90% after the EM-FGA, indicating that the EM-FGA method effectively reduces organic surface contamination.

**Atomic Force Microscopy (AFM):** MoS$_2$ monolayers were also characterized using AFM imaging before and after the EM-FGA to assess changes in surface morphology as illustrated in Figure 6. The topography of the underlying SiO$_2$ substrate changed visibly after the EM-FGA procedure (Figure 6a and 6b), but changes in the MoS$_2$ monolayer were more subtle. The average surface roughness measured on the SiO$_2$ substrate from the topography image reduced

≈ 40 % after the EM-FGA process from 892 pm to 510 pm. On the other hand, surface roughness appeared virtually unchanged for MoS$_2$. Complementary information was obtained by a careful analysis of the phase image of the same sample as seen in Figures 6c and 6d. In those images, we observed distinct and visible changes to both the SiO$_2$ and MoS$_2$ surfaces following EM-FGA. Prior to EM-FGA, the distribution of phase angles (Figure 6c, *inset*) on both the SiO$_2$ and MoS$_2$ surfaces was long-tailed indicating phase non-uniformity. This behavior was consistent with previous observations of contamination in graphene.[70] Following EM-FGA, the phase angle distributions were considerably more uniform and followed a Gaussian distribution suggesting removal of surface contamination. The result agrees with the XPS measurements that show both removal of organic contamination and improvement to the stoichiometry of the MoS$_2$.

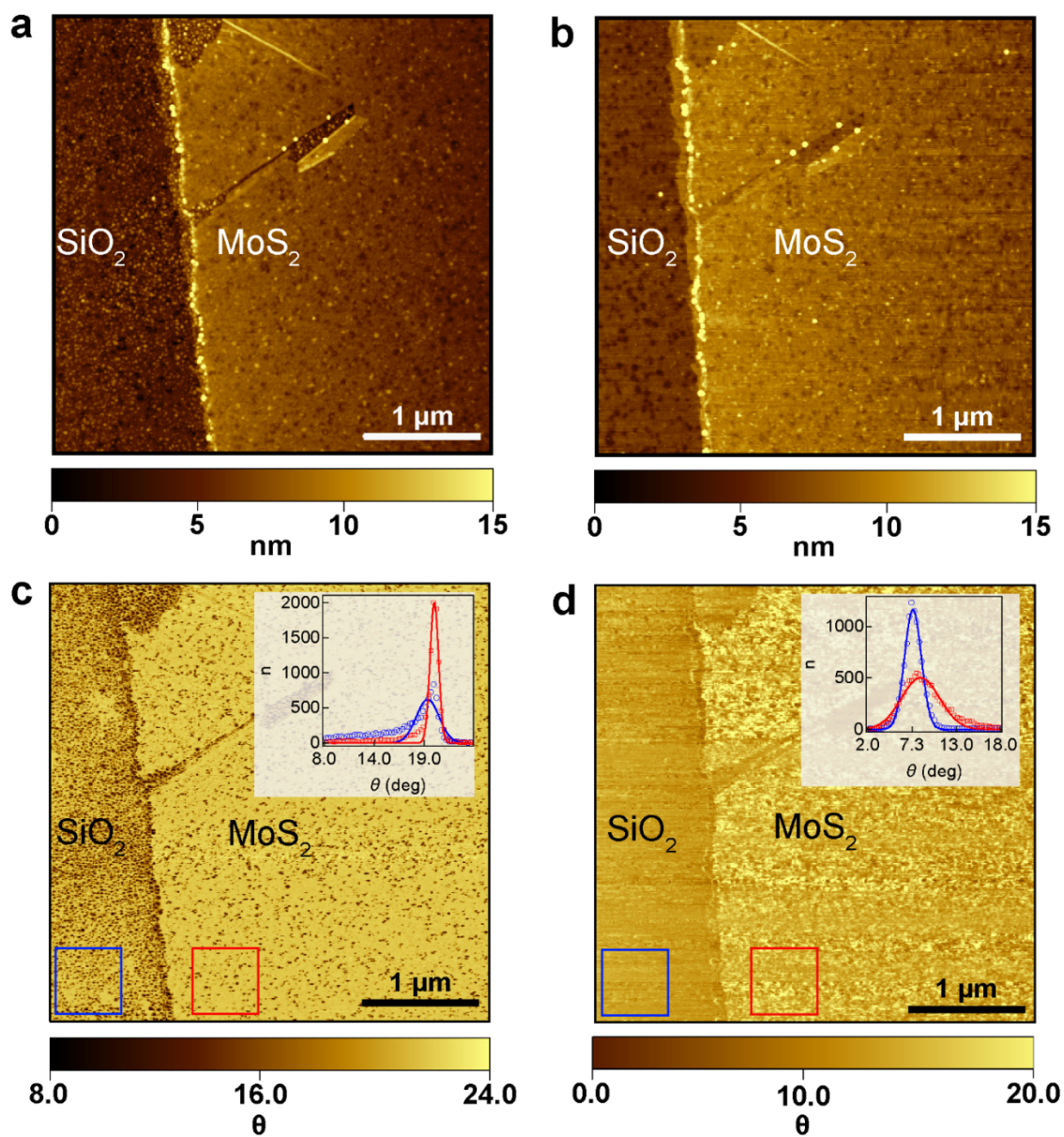

**Figure 6**. AFM images of a MoS$_2$ monolayer before and after exposed material forming gas annealing (EM-FGA). (a) Topographical image of a MoS$_2$ monolayer on SiO$_2$ before EM-FGA. (b) Topographical image of the MoS$_2$ monolayer from (a) on SiO$_2$ after EM-FGA. (c) Phase image of the MoS$_2$ monolayer from (a) on SiO$_2$. *Inset*: distributions of phase shift angles in the selected areas of the SiO$_2$ substrate (*blue*) and MoS$_2$ (*red*). (d) Phase image of the MoS$_2$ monolayer on SiO$_2$ from (b). *Inset*: distributions of phase shift angles in the selected areas of the SiO$_2$ substrate (*blue*) and MoS$_2$ (*red*).

## 3. Conclusion

We demonstrate a new process that markedly improved reproducibility and performance of FETs fabricated from MoS$_2$ monolayers sourced from metal-mediated exfoliation. The EM-FGA process demonstrated in this work improved both the top and back-gate performance of the FETs, as quantified by nearly ideal and reproducible threshold voltages, and Ohmic behavior of the source and drain contacts. Furthermore, common device metrics to estimate performance such as subthreshold slope, drive current, and field-effect mobility of the semiconducting MoS$_2$ were found to be comparable to previous reports of state-of-the-art FETs fabricated by mechanical exfoliation of MoS$_2$. These improvements demonstrate that the EM-FGA remarkably improve the MoS$_2$-gate oxide interfaces by removing trapped charges that can degrade electrical performance. As large area 2D material flakes become more commonplace due to continued interest in the metal-mediated exfoliation method,[42, 67] the improved processing techniques reported here will be critical to enable the fabrication of components from 2D materials in logic circuits for numerous applications.

The combination of the Raman, XPS, and AFM results support the conclusion that the EM-FGA improves the quality and composition of the MoS$_2$ monolayer resulting in improved FET performance. The improvements were found to be two-fold; (i) the EM-FGA process drastically decreased organic contaminants on the semiconducting material and surrounding back-gate dielectric, which can dope the channel and lead to an uncontrolled flatband voltage shift, and (ii) the EM-FGA process eliminated the presence of MoO$_2$ species which can be disadvantageously insulating. Lastly, the increase of anneal time performed on the exposed MoS$_2$ resulted in no observable detrimental effects on FET performance or destruction of the MoS$_2$.

The methods detailed in this work will have an immediate impact when realizing devices that use 2D materials sourced from metal-mediated exfoliation and could help in the development of 2D heterostructure devices where there is a stringent requirement for interface cleanliness and material quality. Finally, EM-FGA can potentially be applied to CVD or PVD grown material to improve material composition. We will extend our processing approach to improve the yield and reproducibility for synthesized 2D materials in future works.

## 4. Methods

**FET Fabrication:** Low resistivity (R < 0.005 Ω-cm) Si wafers with 70 nm $SiO_2$ were cleaned for 15 minutes at 75 °C in an agitated bath of 5:1:1 DI water/ammonium hydroxide/hydrogen peroxide. $MoS_2$ was prepared by gold-mediated exfoliation as described previously[42]: $MoS_2$ was exfoliated from a bulk source onto adhesive tape which was then coated with 110 nm gold Au using electron beam deposition. Thermal-release tape was then used to transfer the gold-coated $MoS_2$ onto the wafers which were subsequently treated with oxygen plasma at 150 W and 30 $cm^3$/min at standard temperature and pressure (STP), 0 °C and 101 kPa, respectively (30 sccm), and 4 Pa (30 mTorr) for 4 minutes to remove residual contamination from the tape while the Au protected the $MoS_2$. The Au was finally removed with Au etchant TFA (8 wt % Iodine, 21 wt % Potassium Iodide, 71 wt % water; Transene Inc., Danvers, MA) for 4 minutes and then cleaned with distilled (DI) water for 10 minutes, acetone for 30 minutes at 45 °C, and then rinsed with DI water and gently blown dry with $N_2$.

After the transfer was complete, the presence of monolayers was confirmed with Raman spectroscopy. Source and drain contacts were patterned onto the entire wafer (i.e., not targeting specific monolayers) by using optical lithography with a stepper. The source and drain contacts were metallized with electron beam deposition of 2 nm Ti and 80 nm Au. Arrays were inspected for source and drain contact overlap of monolayers using optical microscopy and targeted for channel patterning. 5 μm × 5 μm $MoS_2$ channels were patterned using optical lithography and then etched into the monolayer with $XeF_2$ at 100 Pa (1 Torr) and 3 second pulses. For most monolayers, between 10 and 14 pulses were used to fully etch the monolayer. For Raman, XPS, and AFM analysis, samples were processed identically (except for electron beam metal deposition) to mimic processing conditions prior to the EM-FGA.

The EM-FGA was performed on the FETs for 24 hours in a tube furnace at 350 Pa (2.6 Torr) and 400 °C with 100 cm$^3$/min forming gas at standard temperature and pressure (STP), 0 °C and 101 kPa, respectively (100 sccm) of 95:5 $N_2/H_2$. EM-FGA FETs were immediately transferred to a reactor for atomic layer deposition (ALD) of $Al_2O_3$ as opposed to control FETs where the FGA was performed after $Al_2O_3$ deposition. For ALD, saturating doses of trimethylaluminum and water vapor were alternately injected into a custom, warm-walled ALD reactor with a constant flow of ultra-high purity $N_2$ serving as a carrier gas for the reactants and as a purge gas between injections. The substrate was heated to 210 °C while the walls and gas lines were maintained at 110 °C. Under similar conditions, the deposition rate of $Al_2O_3$ was previously found using spectroscopic ellipsometry to be (0.103 ± 0.007) nm per cycle on $SiO_2$. A total of 200 cycles were performed to deposit ≈ 20 nm of top-gate $Al_2O_3$.

Finally, top-gates were patterned onto both sets of FETs using optical lithography and electron beam deposition to deposit 10 nm Ti and 100 nm Au. A second FGA was then performed for 4 hours in a tube furnace at 350 Pa (2.6 Torr) and 400 °C using 100 sccm of 95:5 $N_2/H_2$.

**FET Performance Characterization:** *I-V* characterization was performed during processing using a probe station and parameter analyzer. FETs were tested using standard $I_{DS}V_{DS}$ and $I_{DS}V_G$ measurement protocols for the back-gate where for $I_{DS}V_{DS}$, $V_{DS}$ was swept from 0 V to 1 V and $V_G$ was stepped three times from −10 V, 0 V, and 10 V, and for $I_{DS}V_G$, $V_{DS}$ was stepped six times (0.05, 0.25, 0.45, 0.65, 0.85, 1.05) V and $V_G$ was swept between −30 V to 25 V. A similar protocol was used for the top-gate, but for $I_{DS}V_{DS}$, $V_G$ was set to either −4 V, 0 V, and 4 V, and for $I_{DS}V_G$, $V_G$ was swept from either −6 V to 5 V. All back-gate measurements were made before deposition of the top-gate metal and all top-gate measurements were made with $V_{BG}$ grounded.

The ideal $V_T$ for a monolayer FET was calculated using the method outlined by Ma, et al.[60] where the local channel electrostatic potential ($V_{ch}$) and channel electron density ($n_{ch}$) must satisfy

$$C_q = q^2 g_{2D}\left[1 + \frac{Exp(\frac{E_g}{2k_BT})}{2\cosh(\frac{qV_{ch}}{k_BT})}\right]^{-1}, \tag{1}$$

$$V_G = V_0 + V_{thermal}\ln\left(Exp\left(\frac{n_{ch}}{g_{2D}k_BT}\right) - 1\right) + V_{ox}, \tag{2}$$

$$V_{ch} = V_0 + V_{thermal}\ln\left(Exp\left(\frac{n_{ch}}{g_{2D}k_BT}\right) - 1\right), \tag{3}$$

where $q$ is the elementary charge, $g_{2D}$ is the 2D density of states within the channel, $E_g$ is the band gap energy of MoS$_2$, $k_B$ is the Boltzmann constant, $T$ is the temperature, $V_0 = \frac{E_g}{2q}$, $V_{thermal} = \frac{k_BT}{q}$ is the thermal voltage, and $V_{ox} = \frac{q\, n_{ch}}{C_{ox}}$ is the voltage drop across the gate oxide. Equations (1) – (3) were solved numerically to obtain the ideal $V_T$ for both a back-gate and top-gate interfaces with the monolayer MoS$_2$ FET.

Contact resistance was obtained using 4-point and 2-point probe measurements with the back-gate and top-gates floating. For the 4-point measurement, a constant $V_{DS}$ = 1 V was applied to the first contact, $V$ was measured across the second and third contacts where $I$ was kept constant at zero, and the fourth contact was grounded to yield $R_{14,23} = \frac{V_{23}}{I_{14}} = R_{channel}$. For the 2-point measurement, a constant $V_{DS}$= 1 V was applied between the second and third contacts to yield $R_{23,23} = \frac{V_{23}}{I_{23}} = R_{channel} + 2\,R_{contact}$. Rearranging these two equations for $R_{contact}$ ($R_c$) yields $R_{contact} = \frac{R_{23,23} - R_{14,23}}{2}$. $R_{23,23}$ and $R_{14, 23}$ were found by taking the inverse of the slope of the best fit lines to the *I-V* data as shown in Figure S8.

**Raman Spectroscopy:** Raman spectra were acquired in a Renishaw InVia microscope spectrometer with laser excitation at 514 nm. All Raman peaks were calibrated based on the Si peak (520.7 cm$^{-1}$) and fitted with Gaussian-Lorentzian line shapes to determine the peak position, the line width, and the intensity of different components.

**X-ray Photoelectron Spectroscopy:** XPS Spectra were acquired on a Kratos Axis UltraDLD XPS/UPS system, under a base pressure of 0.135 µPa (10$^{-9}$ Torr), using the monochromatic Al Kα line. The XPS spectra were calibrated using adventitious carbon at ≈ 284.8 eV.

**Atomic Force Microscopy:** AFM images were acquired on an Asylum AFM with the tip in tapping mode to acquire both topographical and phase changes of the MoS$_2$ and the surrounding SiO$_2$ substrate. Scanning was performed at room temperature with settings optimized for 2D materials.

**Supporting Information**

Raman spectra for monolayer, bilayer, and bulk MoS$_2$; field effect transistor (FET) transfer curves for varied exposed material forming gas anneal (EM-FGA) conditions; comparison of transfer curves for control, EM-FGA, and as-exfoliated FETs; complete set of back-gate transfer curves for EM-FGA and control FETs; complete set of back-gate $I_{DS}$-$V_{DS}$ curves for EM-FGA and control FETs; 2-point and 4-point probe *I-V* curves for contact resistance calculations;

complete set of top-gate transfer curves for EM-FGA and control FETs; complete set of top-gate $I_{DS}$-$V_{DS}$ curves for EM-FGA and control FETs; 2-point resistances for EM-FGA FETs

**Acknowledgements**

We would like to thank Ronald Dixson and George Orji for their help interpreting AFM data. N.B.G. and J.B.K. acknowledge support by the National Institute of Standards and Technology (NIST) grant 70NAHB15H023. S.T.L. acknowledges support by the NIST grant 70NANB16H170. Research performed in part at the NIST Center for Nanoscale Science and Technology nanofabrication facility. Certain commercial equipment, instruments, or materials are identified in this paper in order to specify the experimental procedure adequately. Such identification is not intended to imply recommendation or endorsement by the National Institute of Standards and Technology, nor is it intended to imply that the materials or equipment identified are necessarily the best available for the purpose.

Table of Contents (TOC) Graphic

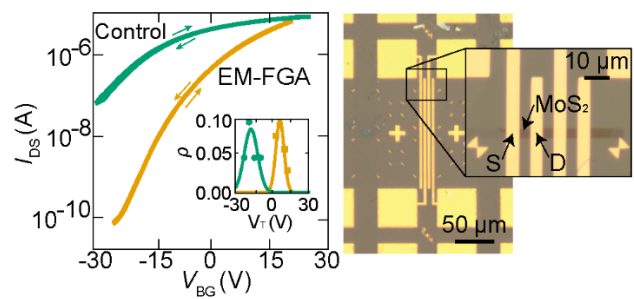

**TOC Image**